\documentclass{article}
\usepackage[dvips]{graphicx}

\begin{document}

\title{Double beta decay experiments}

\author{A.S.~Barabash\thanks{e-mail: barabash@itep.ru} \\
 {\small Institute of Theoretical and Experimental Physics, B.~Cheremushkinskaya 25,} \\
 {\small 117259 Moscow, Russia}}


\date{ }

\maketitle

\begin{abstract}
The present status of double beta decay
experiments
 are reviewed. The results of the most sensitive experiments,
 NEMO-3 and CUORICINO,
 are discussed. Proposals for future double beta decay experiments
are considered. In these experiments sensitivity for the effective neutrino
mass will be on the level of (0.1-0.01) eV.
\end{abstract}

\section{Introduction}

The current interest in neutrinoless double beta decay,
$2\beta(0\nu)$ decay, is that the existence of this process is
closely related to the following fundamental aspects of elementary-
particle
physics \cite{KLA98,FAE01,VER02}: (i) lepton-number
nonconservation, (ii) the
presence of a neutrino mass and its origin, (iii) the existence of
right-handed
currents in electroweak interactions, (iv) the existence of the
Majoron, (v) the
structure of Higgs sector, (vi) supersymmetry, (vii) the existence
of leptoquarks,
(viii) the existence of a heavy sterile neutrino, and (ix) the
existence of a
composite neutrino.

All of these issues are beyond the standard model of electroweak
interaction,
therefore the detection
of $2\beta(0\nu)$ decay would imply the discovery of new physics.
Of course,
interest in this process is caused primarily by the problem of a
neutrino
mass. If $2\beta(0\nu)$ decay is discovered, then according to
current
thinking, this will automatically mean that the rest mass of at
least
one neutrino flavor is nonzero and is of Majorana origin.

Interest in neutrinoless double beta decay has seen a significant
rebirth
in recent years after evidence for neutrino oscillations was
obtained
from the results of atmospheric \cite{link1}  and solar
\cite{link2,link3,link4,link5,link6} neutrino experiments (see,
for example, the
discussions in \cite{link7,link8,link9}).

  This observation of oscillations was recently confirmed by the
KamLAND experiment with reactor
antineutrinos \cite{link10,ARA05} and by the new SNO result
\cite{link11}. These
results are an impressive proof
that neutrinos have a nonzero mass. However, the experiments
studying neutrino oscillations are not sensitive to the nature of
the neutrino mass
(Dirac or Majorana?) and provide no information on the absolute
scale of the neutrino
masses, since such experiments are sensitive only to the $\Delta m^{2}$. 
The detection
and study of $2\beta0\nu$ decay may clarify the following problems
of neutrino physics
(see discussions in \cite{link12,link13,link14}): (i) neutrino
nature; is the neutrino a
Dirac or a Majorana particle?, (ii) absolute neutrino mass scale
(a measurement or a limit
on m$_1$), (iii) the type of neutrino mass hierarchy (normal,
inverted, or  quasidegenerate),
 (iv) CP violation in the lepton sector (measurement of the
Majorana CP-violating phases).

Let us consider three main modes of $2\beta$ decay \footnote{The
decay modes also include
(A, Z) - (A,Z - 2) processes via (i) the emission of two positrons
(2$\beta^{+}$
processes), (ii) the emission of one positron accompanied
by electron capture (EC$\beta^{+}$ processes), and (iii) the
capture of two
orbital electrons (ECEC). For the sake of simplicity, 
we will consider 2$\beta^{-}$ decay. In each case where it will be 
desirable to invoke $2\beta^{+}$ , EC$\beta^{+}$, or ECEC processes, 
this will be 
indicated specifically.}:

\begin{equation}
(A,Z) \rightarrow (A,Z+2) + 2e^{-} + 2\tilde \nu
\end{equation}

\begin{equation}
(A,Z) \rightarrow (A,Z+2) + 2e^{-}
\end{equation}

\begin{equation}
(A,Z) \rightarrow (A,Z+2) + 2e^{-} + \chi^{0}(+ \chi^{0})
\end{equation}

\underline{ $2\beta(2\nu)$ decay (process (1))} is a second-order
process, which is not forbidden
by any conservation law. The detection of this process furnishes
information about nuclear
matrix elements (NME) for $2\nu$ transitions, and this makes it
possible to test the existing
models for calculating these NMEs and contributes to obtaining
deeper insight into the
nuclear-physics aspect of the problem of double beta decay. It is
expected that the accumulation
of experimental information about $2\beta(2\nu)$ processes will
improve the quality
of the calculations of NMEs, both for $2\nu$ and for $0\nu$ decay.
Moreover,
the study can yield a careful
investigation of the
time dependence of the coupling constant for weak interactions
\cite{BAR98,BAR00}.

\underline{ $2\beta(0\nu)$ decay (process (2))} violates the law
of lepton-number conservation
($\Delta L =2$)
and requires that the Majorana neutrino has a nonzero rest mass or
that an admixture of
right-handed currents be present in weak interaction. Also, this
process is possible in some
supersymmetric models, where $2\beta(0\nu)$ decay is initiated by
the exchange of supersymmetric
particles. This decay also arises in models featuring an extended
Higgs sector within
electroweak-interaction theory and in some other cases
\cite{KLA98}.

\underline{ $2\beta(0\nu\chi^{0})$ decay (process (3))} requires
the existence of a Majoron - it is a massless
Goldstone boson that arises upon a global breakdown of (B -L)
symmetry, where B and L are,
respectively, the baryon and the lepton number. The Majoron, if it
exists, could play a significant role
in the history of the early Universe and in the evolution of
stars. The model of a triplet
Majoron \cite{GEL81} was disproved in 1989
by the data on the decay width of the $Z^{0}$ boson that were
obtained at the LEP accelerator (CERN,
Switzerland). Despite this, some new models were proposed
\cite{MOH91,BER92}, where $2\beta(0\nu\chi^{0})$
decay is possible
and where there are no contradictions with the LEP data. A
$2\beta$-decay model that involves the
emission of two Majorons was proposed within supersymmetric
theories \cite{MOH88} and several other models of the
Majoron were proposed in the 1990s. By the term "Majoron," one
means here massless or light bosons
that are associated with neutrinos. In these models, the Majoron
can carry a lepton charge and is
not required to be a Goldstone boson \cite{BUR93}. A decay process
that involves the emission of two "Majorons"
is also possible \cite{BAM95}. In models featuring a vector
Majoron, the Majoron is the longitudinal
component of a massive gauge boson emitted in 2$\beta$ decay
\cite{CAR93}. For the sake of simplicity, each such
object is referred to here as a Majoron.
In the Ref. \cite{MOH00}, a "bulk" Majoron model was proposed
in the context of the "brane-bulk" scenario for particle physics.

\begin{figure*}
\begin{center}
\resizebox{0.5\textwidth}{!}{\includegraphics{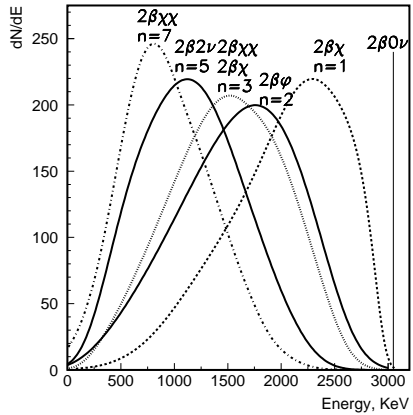}}
\caption{Energy spectra of different modes of $2\beta2\nu$
$(n=5)$,
$2\beta\chi^{0}$ $(n=1~$,$~2$ and$~3)$ and
$2\beta\chi^{0}\chi^{0} (n=3~$and$~7)$ decays of $~^{100}$Mo.}
\label{fig_figure1}
\end{center}
\end{figure*}

The possible two electrons energy spectra for different 2$\beta$
decay modes of $^{100}$Mo are shown
in Fig ~\ref{fig_figure1}. Here n is spectral 
index, which defines
the shape of the spectrum. For example, for ordinary Majoron n = 1, 
for 2$\nu$ decay n = 5, in case of bulk Majoron n = 2 
and for the process with two Majoron emission n= 3 or 7.

\begin{figure*}
\begin{center}
\resizebox{0.5\textwidth}{!}{\includegraphics{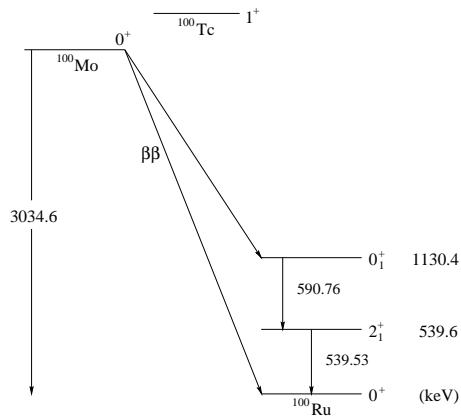}}
\caption{Levels scheme for $^{100}$Mo - $^{100}$Tc - $^{100}$Ru.}
\label{fig_figure2}
\end{center}
\end{figure*}

\section{Results of experimental investigations }

The number of possible candidates for double-beta decay is quite
large - there are
approximately 30 nuclei.\footnote{Approximately the same number of
nuclei can undergo double electron
capture, while twenty nuclei and six nuclei can undergo,
respectively, EC$\beta^{+}$ and 2$\beta^{+}$ decay (see the tables
in \cite{TRE02}).

} However, nuclei for which the double-beta-transition energy
$E_{2\beta}$ is in excess
of 2 MeV are of greatest interest, since the double-beta-decay
probability strongly depends on the transition
energy. In transitions to excited states of the daughter nucleus,
the excitation energy is removed via the
emission of one or two photons, which can be detected, and this
can therefore serve as an additional source
of information about double-beta decay. Fig
~\ref{fig_figure2} shows the diagram of energy levels in the
$^{100}$Mo - $^{100}$Tc - $^{100}$Ru nuclear triplet (as example).

\subsection{Two neutrino double beta decay}\label{subsec:general}

This decay was first recorded in 1949 in a geochemical experiment
with $^{130}$Te \cite{ING49}; in 1967,
it was also found for $^{82}$Se \cite{KIR67}.
Attempts to observe this decay in a direct experiment employing
counters had been futile for
a long time. Only in 1987 could M. Moe, who used a time-projection
chamber (TPC), observe $2\beta(2\nu)$ decay
in $^{82}$Se
for the first time \cite{ELL87}. Within the next few years,
experiments employing counters were able to detect
$2\beta(2\nu)$
decay in many nuclei. In $^{100}$Mo \cite{BAR95,BAR99,BRA01}, and
$^{150}$Nd \cite{BAR04}
$2\beta(2\nu)$ decay to the $0^{+}$ excited state of the
daughter nucleus was recorded. Also, the
$2\beta(2\nu)$ decay of $^{238}$U was detected in a radiochemical
experiment \cite{TUR91}, and in a geochemical experiment for the
first time the ECEC process was detected in $^{130}$Ba
\cite{MES01}. Table 1 displays the present-day averaged and
recommended values of
T$_{1/2}$(2$\nu$) from \cite{BAR05}.
At the present-time, experiments devoted to detecting
$2\beta(2\nu)$ decay are approaching a new level
where it is insufficient to restrict oneself to recording the
decay process, but it is necessary to
measure numerous parameters of this process to a high precision.
Tracking detectors that are able to record both
the energy of each electron and the angle at which they diverge
are the most appropriate instruments for
solving this problem.

\begin{table}[ht]
\label{Table1}
\caption{Average and recommended $T_{1/2}(2\nu)$ values (from
\cite{BAR05}).}
\vspace{0.5cm}
\begin{center}
\begin{tabular}{cc}
\hline
Isotope & $T_{1/2}(2\nu)$ \\
\hline
$^{48}$Ca & $4.3^{+2.1}_{-1.0}\cdot10^{19}$ \\
$^{76}$Ge & $(1.5 \pm 0.1)\cdot10^{21}$ \\
$^{82}$Se & $(0.92 \pm 0.07)\cdot10^{20}$ \\
$^{96}$Zr & $(2.0 \pm 0.3)\cdot10^{19}$ \\
$^{100}$Mo & $(7.1 \pm 0.4)\cdot10^{18}$ \\
$^{100}$Mo-$^{100}$Ru$(0^{+}_{1})$ & $(6.8 \pm 1.2)\cdot10^{20}$ \\
$^{116}$Cd & $(3.0 \pm 0.2)\cdot10^{19}$\\
$^{128}$Te & $(2.5 \pm 0.3)\cdot10^{24}$ \\
$^{130}$Te & $(0.9 \pm 0.1)\cdot10^{21}$ \\
$^{150}$Nd & $(7.8 \pm 0.7)\cdot10^{18}$ \\
$^{150}$Nd-$^{150}$Sm$(0^{+}_{1})$ & $1.4^{+0.5}_{-0.4}\cdot10^{20}$
\\
$^{238}$U & $(2.0 \pm 0.6)\cdot10^{21}$  \\
$^{130}$Ba; ECEC(2$\nu$) & $(2.2 \pm 0.5)\cdot10^{21}$  \\
\hline
\end{tabular}
\end{center}
\end{table}

\subsection{Neutrinoless double beta decay}

In contrast to two-neutrino decay, neutrinoless double-beta decay
has not yet been observed
\footnote{The possible
exception is the result with $^{76}$Ge, published by a fraction of
the  Heidelberg-Moscow
Collaboration, $T_{1/2} = 1.2\cdot 10^{25}$ y \cite{KLA04} (see
Table 2). The Moscow portion of the
Collaboration does not agree with this conclusion \cite{BAK03} and
there are others who are critical
of this result \cite{STR05}. Thus at the present time this
"positive" result is not accepted by the
"2$\beta$ decay community" and it has to be checked by new
experiments.}, although
from the experimental point of view, it is easier to detect it. In
this case, one seeks, in the
experimental spectrum, a peak of energy equal to the double-beta-
transition energy and of width determined
by the detector's resolution.

The present-day constraints on the existence of
$2\beta(0\nu)$ decay are presented in Table 2 for the nuclei
that are the most promising candidates. In calculating constraints
on $\langle m_{\nu} \rangle$, the
nuclear matrix elements from \cite{SIM99,STO01,CIV03} were used (3-d column). It
is advisable to employ the calculations from
these studies, because the calculations are the most thorough and
take into account the most recent theoretical
achievements. The respective phase-space volumes were taken from \cite{SUH98}. 
In column four, limits on $\langle m_{\nu} \rangle$,
which were obtained using the NMEs
from a recent paper \cite{ROD05}. In this paper $g_{pp}$ values ($g_{pp}$ is 
parameter of QRPA theory) were fixed using experimental 
half-life values for $2\nu$ decay and then
NME(0$\nu$) were calculated. 
Those authors analyze results of all existing QRPA calculations and 
demonstrate that their approach
gives most accurate and reliable values for NMEs (but see, nevertheless, 
critics in \cite{SUH05}).

One can see from the Table 1 that in framework of NME calculations 
from \cite{SIM99,STO01,CIV03} the limits on 
$\langle m_{\nu} \rangle$ for $^{130}$Te and $^{100}$Mo are comparable 
with $^{76}$Ge results. And now one can not 
select any experiment as absolutely best one. From another side exactly 
the assemblage of sensitive experiments
for different nuclei permits to increase reliability of the limit 
on $\langle m_{\nu} \rangle$. Present conservative limit can be set as 0.9 eV.

\begin{table}[ht]
\label{Table2}
\caption{Best present results on $2\beta(0\nu)$ decay (limits at
90\% C.L.). $^{*)}$ Conservative limit from \cite{BER02} is presented. 
$^{**)}$ Current experiments.}
\vspace{0.5cm}
\begin{center}
\begin{tabular}{ccccc}
\hline
Isotope & $T_{1/2}$, y & $\langle m_{\nu} \rangle$, eV & $\langle
m_{\nu} \rangle$, eV & Experiment \\
& & \cite{SIM99,STO01,CIV03} & \cite{ROD05}  \\
\hline
$^{76}$Ge & $>1.9\cdot10^{25}$ & $<0.33-0.84$ & $<0.53-0.59$ & HM
\cite{KLA01} \\
& $\simeq 1.2\cdot10^{25}$(?) & $\simeq 0.5-1.3(?)$ & $\simeq
0.7(?)$ & Part of HM \cite{KLA04} \\
& $>1.6\cdot10^{25}$ & $<0.36-0.92$ & $<0.58-0.64$ & IGEX
\cite{AAL02} \\
\hline
$^{130}$Te & $>1.8\cdot10^{24}$ & $<0.4-0.9$ & $<1-1.6$ &
CUORICINO$^{**)}$ \cite{ARNA05} \\
$^{100}$Mo & $>4.6\cdot10^{23}$ & $<0.65-1.0$ & $<2.4-3.0$ & NEMO-
3$^{**)}$ \cite{ARN05} \\
$^{136}$Xe & $>4.5\cdot10^{23*)}$ & $<0.8-4.7$ & $<2.9-5.6$ & DAMA
\cite{BER02} \\
$^{116}$Cd & $>1.7\cdot10^{23}$ & $<1.4-2.5$ & $<3.7-4.3$ &
SOLOTVINO \cite{DAN03} \\
$^{82}$Se & $>1\cdot10^{23}$ & $<1.7-3.7$ & $<3.8-4.7$ & NEMO-3$^{**)}$
\cite{ARN05} \\
\hline
\end{tabular}
\end{center}
\end{table}

\subsection{Double beta decay with Majoron emission}

Table 3 displays the best present-day constraints for a "ordinary"
(n = 1) Majoron.
The "nonstandard" models of the Majoron were experimentally tested
in \cite{GUN97} for $^{76}$Ge
and in \cite{ARN00} for $^{100}$Mo,
$^{116}$Cd, $^{82}$Se, and $^{96}$Zr. Constraints on the decay
modes involving the emission of two Majorons were also
obtained for $^{100}$Mo \cite{TAN93}, $^{116}$Cd \cite{DAN03}, and
$^{130}$Te \cite{ARN03}. In recent NEMO Collaboration paper \cite{ARN06} 
new results of search for these processes in $^{100}$Mo and $^{82}$Se 
obtained with NEMO-3 detector
were presented. 
Table 5 gives the best experimental
constraints on decays accompanied by the emission of one or two
Majorons (for n = 2, 3, and 7).
\begin{table}[ht]
\label{Table3}
\caption{Best present results on $2\beta(0\nu\chi^{0})$ decay
(ordinary Majoron) at 90\% C.L. The NME from the 
following works were used, 3-d column: \cite{SIM99,STO01,CIV03}, 4-th 
column: \cite{ROD05}. $^{*)}$ Conservative limit from \cite{BER02} is presented.}
\vspace{0.5cm}
\begin{center}
\begin{tabular}{cccc}
\hline
Isotope & $T_{1/2}$, y & $\langle g_{ee} \rangle$, \cite{SIM99,STO01,CIV03}
 & $\langle
g_{ee} \rangle$, \cite{ROD05} \\  \\
\hline
$^{76}$Ge & $>6.4\cdot10^{22}$ \cite{KLA01} & $<(1.2-3.0)\cdot10^{-
4}$ & $<(1.9-2.3)\cdot10^{-4}$ \\
$^{82}$Se & $>1.5\cdot10^{22}$ \cite{ARN06} & $<(0.66-
1.4)\cdot10^{-4}$ & $<(1.2-1.9)\cdot10^{-4}$ \\
$^{100}$Mo & $>2.7\cdot10^{22}$ \cite{ARN06} & $<(0.4-
0.7)\cdot10^{-4}$ & $<(1.6-1.8)\cdot10^{-4}$ \\
$^{116}$Cd & $>8\cdot10^{21}$ \cite{DAN03} & $<(1.0-2.0)\cdot10^{-
4}$ & $<(2.8-3.3)\cdot10^{-4}$ \\
$^{128}$Te & $>2\cdot10^{24}$(geochem)\cite{MAN91} & $<(0.7-
1.6)\cdot10^{-4}$ & $<(1.9-2.4)\cdot10^{-4}$ \\
$^{136}$Xe & $>1.6\cdot10^{22*)}$ \cite{BER02} & $<(0.7-5.0)\cdot10^{-
4}$ & $<(3.4-4.4)\cdot10^{-4}$ \\
\hline
\end{tabular}
\end{center}
\end{table}

Hence at the present time only limits on double beta decay with
Majoron emission have been obtained (see table 3 and 4).
Conservative present limit on the coupling constant of ordinary 
Majoron to the
neutrino is $\langle g_{ee} \rangle < 1.8 \cdot 10^{-4}$
\cite{ARN06}.
\begin{table}[ht]
\label{Table4}
\caption{Best present limits on $T_{1/2}$ for decay with one and
two Majorons at 90\% C.L. for modes
with spectral index n = 2, n = 3 and n = 7.}
\vspace{0.5cm}
\begin{center}
\begin{tabular}{cccc}
\hline
Isotope & n = 2 & n = 3 &  n = 7  \\
\hline
$^{76}$Ge & - & $>5.8\cdot10^{21}$ \cite{GUN97} &
$>6.6\cdot10^{21}$ \cite{GUN97} \\
$^{82}$Se & $>6\cdot10^{21}$ \cite{ARN06} & $>3.1\cdot10^{21}$
\cite{ARN06} & $>5\cdot10^{20}$ \cite{ARN06} \\
$^{96}$Zr & - & $>6.3\cdot10^{19}$ \cite{ARN00} &
$>2.4\cdot10^{19}$ \cite{ARN00} \\
$^{100}$Mo & $>1.7\cdot10^{22}$ \cite{ARN06} & $>1\cdot10^{22}$
\cite{ARN06} & $>7\cdot10^{19}$ \cite{ARN06} \\
$^{116}$Cd & $>1.7\cdot10^{21}$ \cite{DAN03} & $>8\cdot10^{20}$
\cite{DAN03} & $>3.1\cdot10^{19}$ \cite{DAN03} \\
$^{130}$Te & - & $>9\cdot10^{20}$ \cite{ARN03} & - \\
\hline
\end{tabular}
\end{center}
\end{table}

\section{Best present experiments}

In this section the two large-scale current experiments NEMO-3 and
CUORICINO are discussed.

\subsection{NEMO-3 experiment \cite{ARN05,ARN04, ARN05a}}

\begin{figure*}
\begin{center}
\resizebox{0.5\textwidth}{!}{\includegraphics{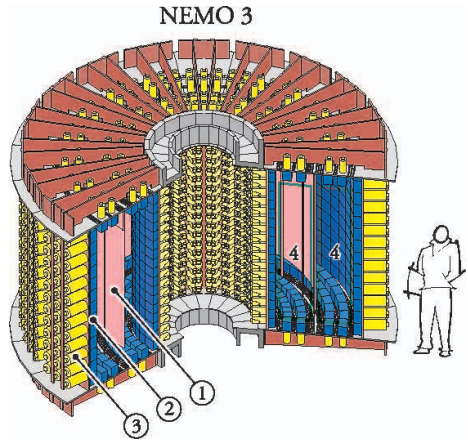}}
\caption{ The NEMO-3 detector without shielding. 1 -- source foil;
2-- plastic scintillator; 3 -- low radioactivity PMT; 4 --
tracking chamber.}
\label{fig_figure3}
\end{center}
\end{figure*}

This is a tracking experiment that, in contrast to experiments
with $^{76}$Ge, detects not only the total energy deposition,
but also other parameters of the process, including the energy of
the individual electrons, angle between them,
and the coordinates of the event in the source plane. The
performance of the detector was studied with the NEMO-2
prototype \cite{ARN95}. Since June of 2002, the NEMO-3 detector
has operated at the Frejus Underground Laboratory (France)
located at a depth of 4800 m w.e. The detector has a cylindrical
structure and consists of 20 identical sectors
 (see Fig ~\ref{fig_figure3}).
A thin (about 30-60 mg/cm$^{2}$) source containing beta-decaying
nuclei and having a total area of 20 m$^{2}$ and a weight
of up
to 10 kg was placed in the detector. The basic principles of
detection are identical to those used in the NEMO-2
detector. The energy of the electrons is measured by plastic
scintillators (1940 individual counters), while the tracks
are reconstructed on the basis of information obtained in the
planes of Geiger cells (6180 cells) surrounding the source
on both sides. The tracking volume of the detector is filled with
a mixture consisting of $\sim$ 95\% He, 4\% alcohol, 1\% Ar
and 0.1\% water at slightly above atmospheric
pressure. In addition, a magnetic field of strength of about 25 G
parallel to the detector axis is created by a solenoid
surrounding the detector. The magnetic field is used to identify
electron-positron pairs and, hence, to suppress this
source of background.

The main characteristics of the detector are the following: the
energy resolution of the scintillation counters lies in
the interval 14-17\% FWHM for electrons of energy 1 MeV; the time
resolution is 250 ps for an electron energy of
1 MeV; and the accuracy in reconstructing of the vertex of 2e$^{-
}$ events is about 1 cm.
The detector is surrounded by a passive shield consisting of 20 cm
of steel and 30 cm of water. The level of
radioactive impurities in structural materials of the detector and
of the passive shield was tested in measurements
with low-background HPGe detectors.

Measurements with the NEMO-3 detector revealed that tracking
information, combined with time and energy measurements,
makes it possible to suppress the background efficiently. That
NEMO-3 can be used to investigate almost all isotopes
of interest is a distinctive feature of this facility. At the
present time, such investigations are being performed
for seven isotopes; these are $^{100}$Mo, $^{82}$Se, $^{116}$Cd,
$^{150}$Nd, $^{96}$Zr, $^{130}$Te, and $^{48}$Ca (see
Table 5). In addition, foils from copper and natural (not
enriched) tellurium are placed in the detector for performing
background measurements.

\begin{table}[ht]
\label{Table5}
\caption{Investigated isotopes with NEMO-3}
\vspace{0.5cm}
\begin{center}
\begin{tabular}{cccccccc}
\hline
Isotope & $^{100}$Mo & $^{82}$Se & $^{130}$Te & $^{116}$Cd &
$^{150}$Nd & $^{96}$Zr & $^{48}$Ca  \\
\hline
Enrichment, & 97 & 97 & 89 & 93 & 91 & 57 & 73 \\
\% & & & & & & & \\
Mass of & 6914 & 932 & 454 & 405 & 36.6 & 9.4 & 7.0 \\
isotope, g & & & & & & & \\
\hline
\end{tabular}
\end{center}
\end{table}

\begin{figure*}
\begin{center}
\includegraphics[scale=0.3]{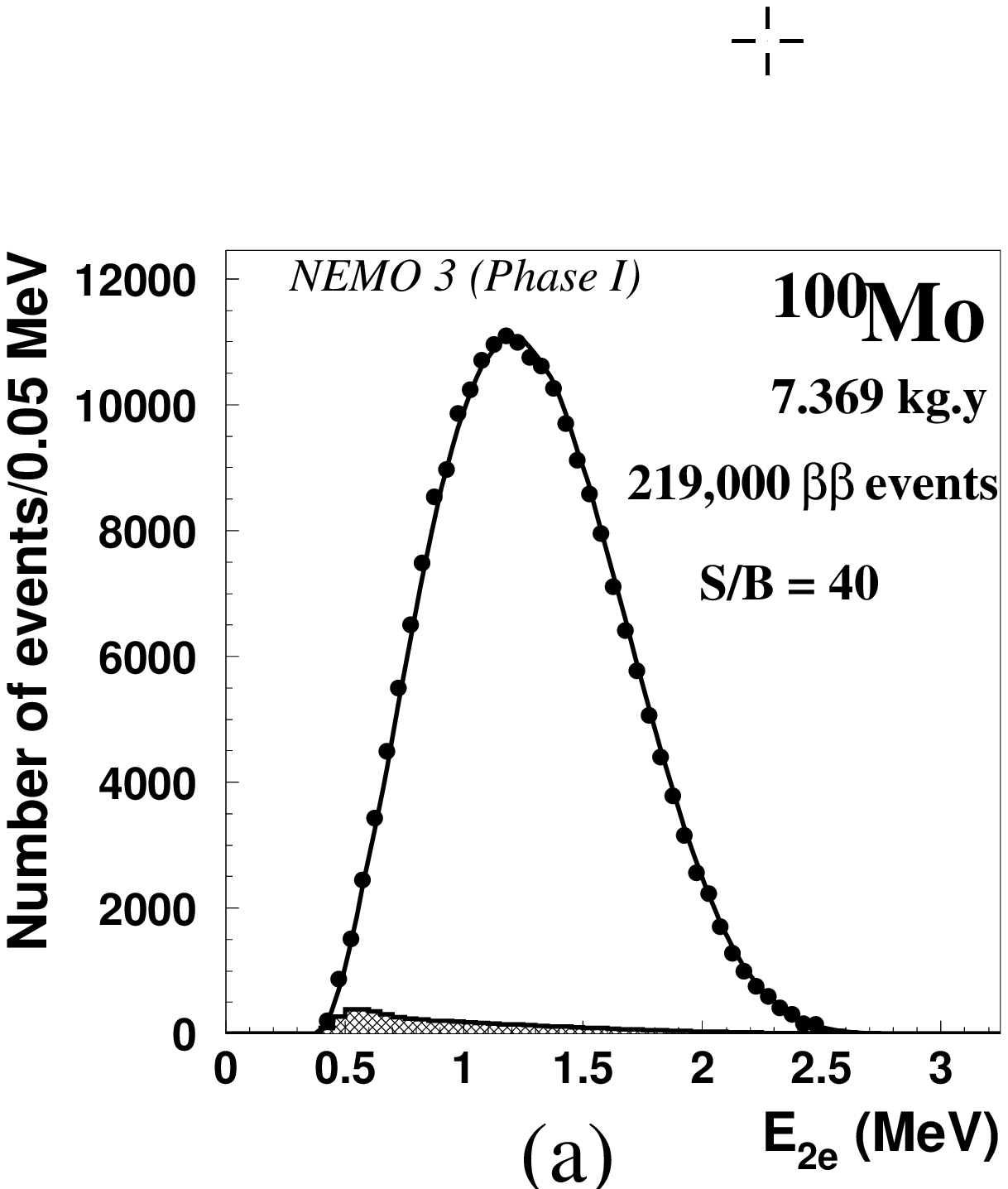}
\includegraphics[scale=0.3]{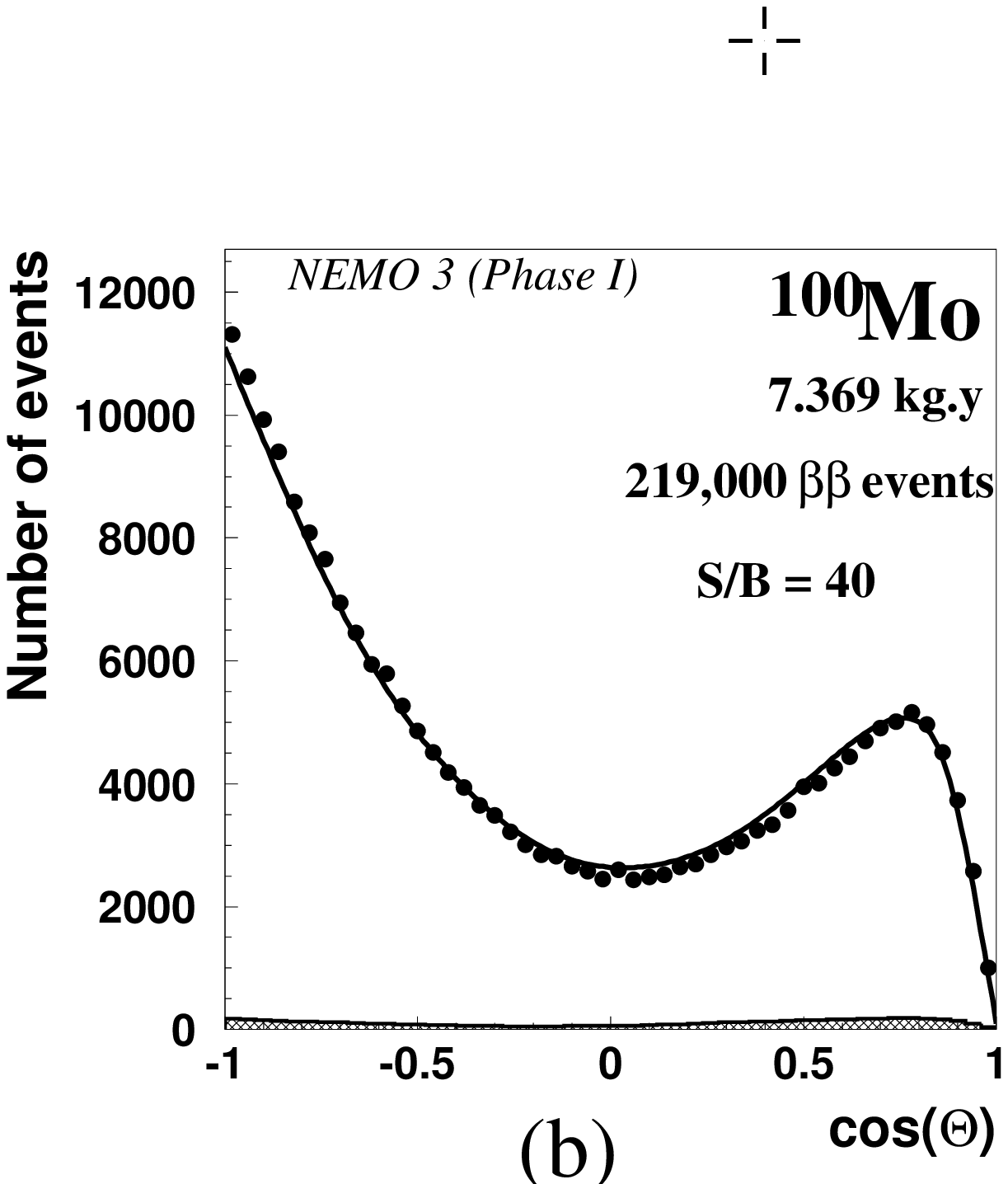}
\includegraphics[scale=0.3]{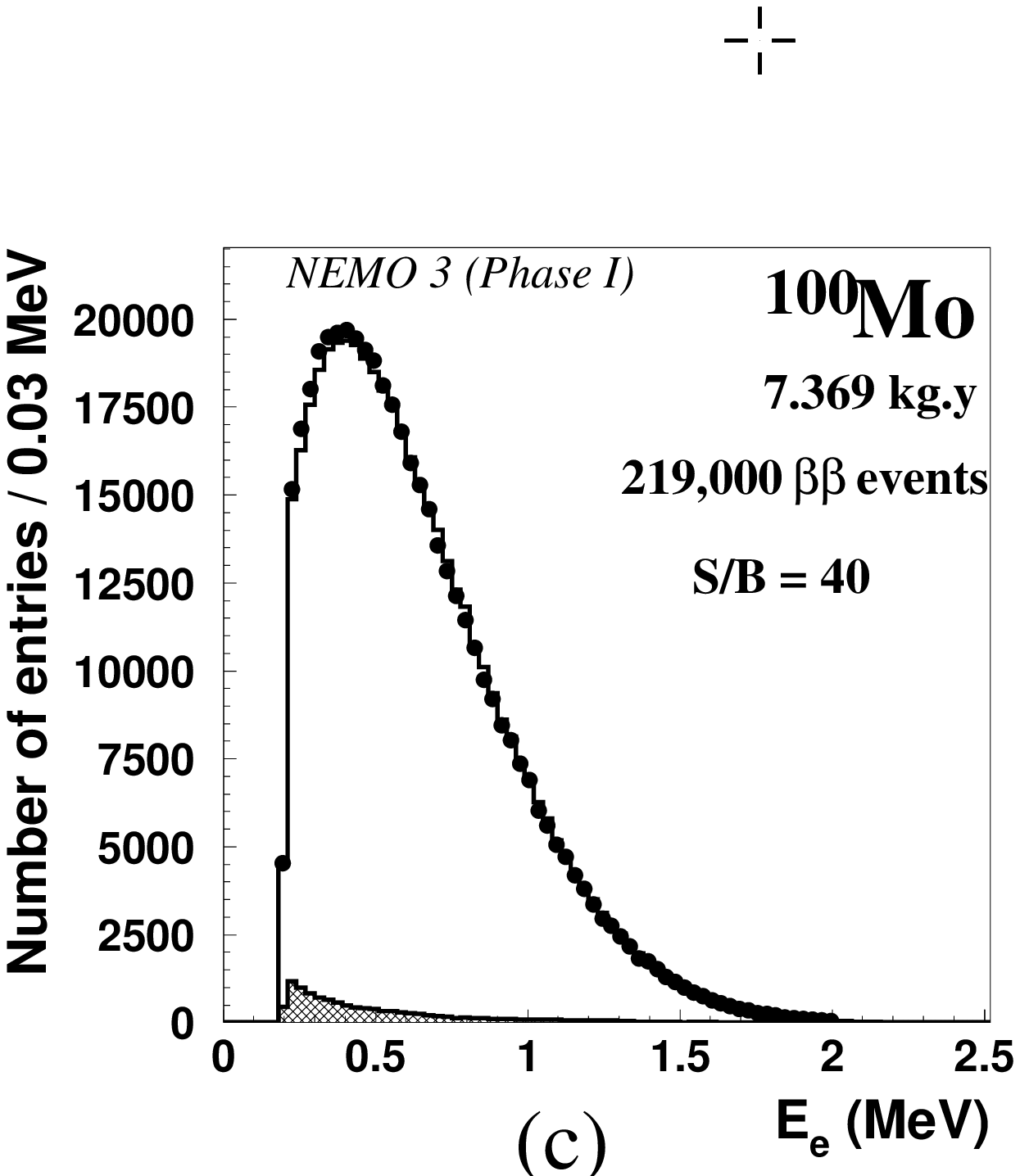}
\caption{\label{fig:figure4} (a) Energy sum spectrum of the two
electrons, (b) angular
distribution of the two electrons and (c) single energy spectrum
of the electrons, after background
subtraction from $^{100}$Mo with 7.369 kg$\cdot$years  
exposure \cite{ARN05}.
The solid line corresponds to the expected
spectrum from $2\beta(2\nu)$ simulations and the shaded
histogram is the subtracted background
computed by Monte-Carlo simulations.}
\end{center}
\end{figure*}

\begin{figure}
\begin{center}
\includegraphics[scale=0.4]{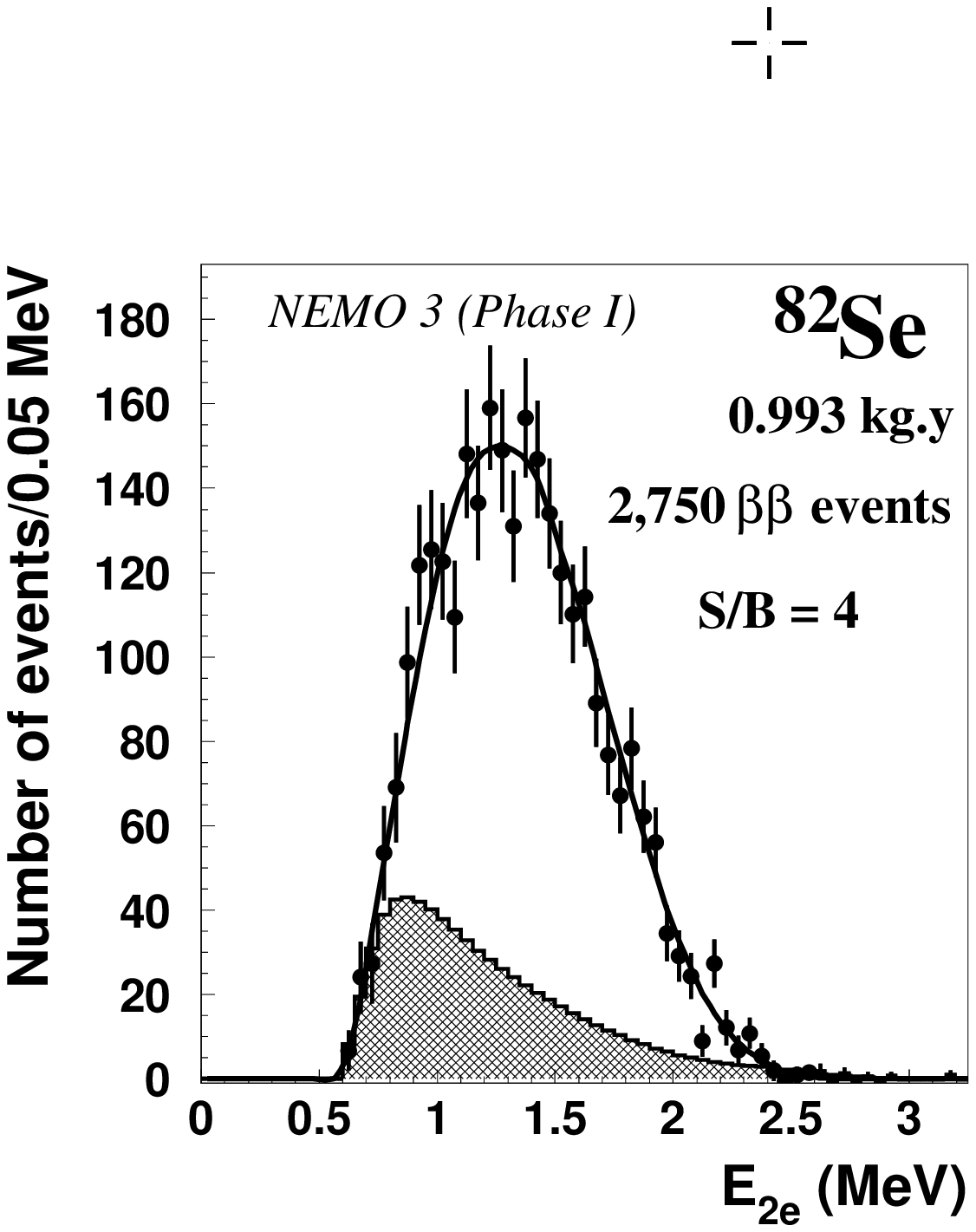}
\caption{\label{fig:figure5} Energy sum spectrum of the two
electrons after background subtraction
from $^{82}$Se with 0.993 kg$\cdot$years exposure (same legend as
Fig.~\ref{fig:figure4}) \cite{ARN05}. The signal contains
 2,750 2$\beta$ events and the signal-to-background ratio is
4.}
\end{center}
\end{figure}

Fig.~\ref{fig:figure4} and Fig.~\ref{fig:figure5}  display the
spectrum of $2\beta(2\nu)$ events,
in $^{100}$Mo and $^{82}$Se that were collected over 389 days
\cite{ARN05}. For $^{100}$Mo angular distribution
(Fig. 4b) and single electron spectrum (Fig. 4c) are also shown.
The total number of useful events is about 219,000 which is much greater
than the total statistics of all of the preceding
experiments with $^{100}$Mo (and even greater then total statistics of all previous 
$2\beta$ decay experiments!). It should also be noted that the
background is as low as about $2.5\%$ of the total number of
useful events. By employing the calculated values of the detection
efficiencies for
$2\beta(2\nu)$ events, the following half-life values were
obtained for $^{100}$Mo and $^{82}$Se:

\begin{equation}
T_{1/2}(^{100}Mo;2\nu) = [7.11 \pm 0.02(stat) \pm 0.54(syst)
]\cdot 10^{18}\; y\
\end{equation}

\begin{equation}
T_{1/2}(^{82}Se;2\nu) = [9.6 \pm 0.3(stat) \pm 1.0(syst) ]\cdot
10^{19}\; y\
\end{equation}

These results and results for $^{116}$Cd, $^{96}$Zr and $^{150}$Nd
are presented in Table 6. Notice that values for
$^{100}$Mo and $^{116}$Cd have been obtained on the assumption
that SSD mechanism is valid \footnote{Validity of SSD mechanism
in $^{100}$Mo was demonstrated using analysis of single electron 
spectrum (see \cite{ARN04}). In the case of $^{116}$Cd this is still 
a hypothesis.} \cite{SIM01,DOM05}.
Systematic uncertainties can be decreased using special
calibrations and can be improved up to $\sim (3-5)\%$.

\begin{table}[ht]
\label{Table6}
\caption{Two neutrino half-life values for different nuclei
obtained in
the NEMO-3 experiment (for $^{116}$Cd, $^{96}$Zr
and $^{150}$Nd results are preliminary). 
First error is statistical and second is systematic. 
S/B is the signal-to-
background ratio.}
\vspace{0.5cm}
\begin{center}
\begin{tabular}{ccccc}
\hline
Isotope & Measurement & Number of & S/B & $T_{1/2}(2\nu)$, y \\
& time, days & $2\nu$ events & &  \\
\hline
$^{100}$Mo & 389 & 219000 & 40 & $(7.11 \pm 0.02 \pm
0.54 )\cdot 10^{18}$  \\
$^{82}$Se & 389 & 2750 & 4 & $(9.6 \pm 0.3 \pm 1.0)
\cdot 10^{19}$  \\
$^{116}$Cd & 168.4 & 1371 & 7.5 & $(2.8 \pm 0.1 \pm
0.3)\cdot 10^{19}$  \\
$^{96}$Zr & 168.4 & 72 & 0.9 & $(2.0 \pm 0.3 \pm 0.2)
\cdot 10^{19}$  \\
$^{150}$Nd & 168.4 & 449 & 2.8 & $(9.7 \pm 0.7 \pm 1.0)
\cdot 10^{18}$  \\
\hline
\end{tabular}
\end{center}
\end{table}

Fig. 6 show the tail of the two-electron energy sum spectrum in
the $2\beta(0\nu)$ energy window
for $^{100}$Mo and $^{82}$Se.
One can see that experimental spectrum is in good agreement with
calculated spectrum, which was obtained taking
into account all sources of background. Using a maximal
likelihood method the following
limits on neutrinoless double beta decay of $^{100}$Mo and
$^{82}$Se (mass mechanism; 90\% C.L.) have been obtained:

\begin{equation}
T_{1/2}(^{100}Mo;0\nu) > 4.6\cdot 10^{23}\; y\
\end{equation}

\begin{equation}
T_{1/2}(^{82}Se;0\nu) > 1\cdot 10^{23}\; y\
\end{equation}

These limits are approximately one order of magnitude better than
results of previous experiments \cite{EJI01,ARN98}.

Using NME values from \cite{SIM99,STO01,CIV03} the bound on
$\langle m_{\nu} \rangle$ is 0.65-1.0 eV for $^{100}$Mo
and 1.7-3.7 eV for $^{82}$Se. If one will use NMEs from
\cite{ROD05}
then $\langle m_{\nu} \rangle < 2.4-3.0$ eV and $< 3.8-4.7$ eV,
respectively.

In this experiment the best present limits on all possible modes
of double beta decay with Majoron emission
have been obtained \cite{ARN06} too (see Tables 3 and 4).

For this first running period (Phase~I) presented here, radon was 
the dominant background in $2\beta(0\nu)$ decay energy region. 
It has now been significantly reduced by a factor $\sim$10 by 
a radon-tight tent enclosing 
the detector and a radon-trap facility in operation since December 
2004 which has 
started a second running period (Phase~II).
After five years of data collection, the expected sensitivity 
at 90\% C.L will be 
$T_{1/2}(\beta\beta0\nu) > 2 \times 10^{24}$~y for $^{100}$Mo and 
$8 \times 10^{23}$~y for $^{82}$Se, corresponding to 
$\langle m_{\nu} \rangle < 0.3-1.4$~eV for $^{100}$Mo and 
$\langle m_{\nu} \rangle < 0.6-1.7$~eV for $^{82}$Se. At the same time 
the search for decay with Majoron emission with record sensitivity and 
precise investigation of $2\beta(2\nu)$ decay in seven mentioned above nuclei 
will be continued.

\begin{figure*}
\begin{center}
\includegraphics[scale=0.2]{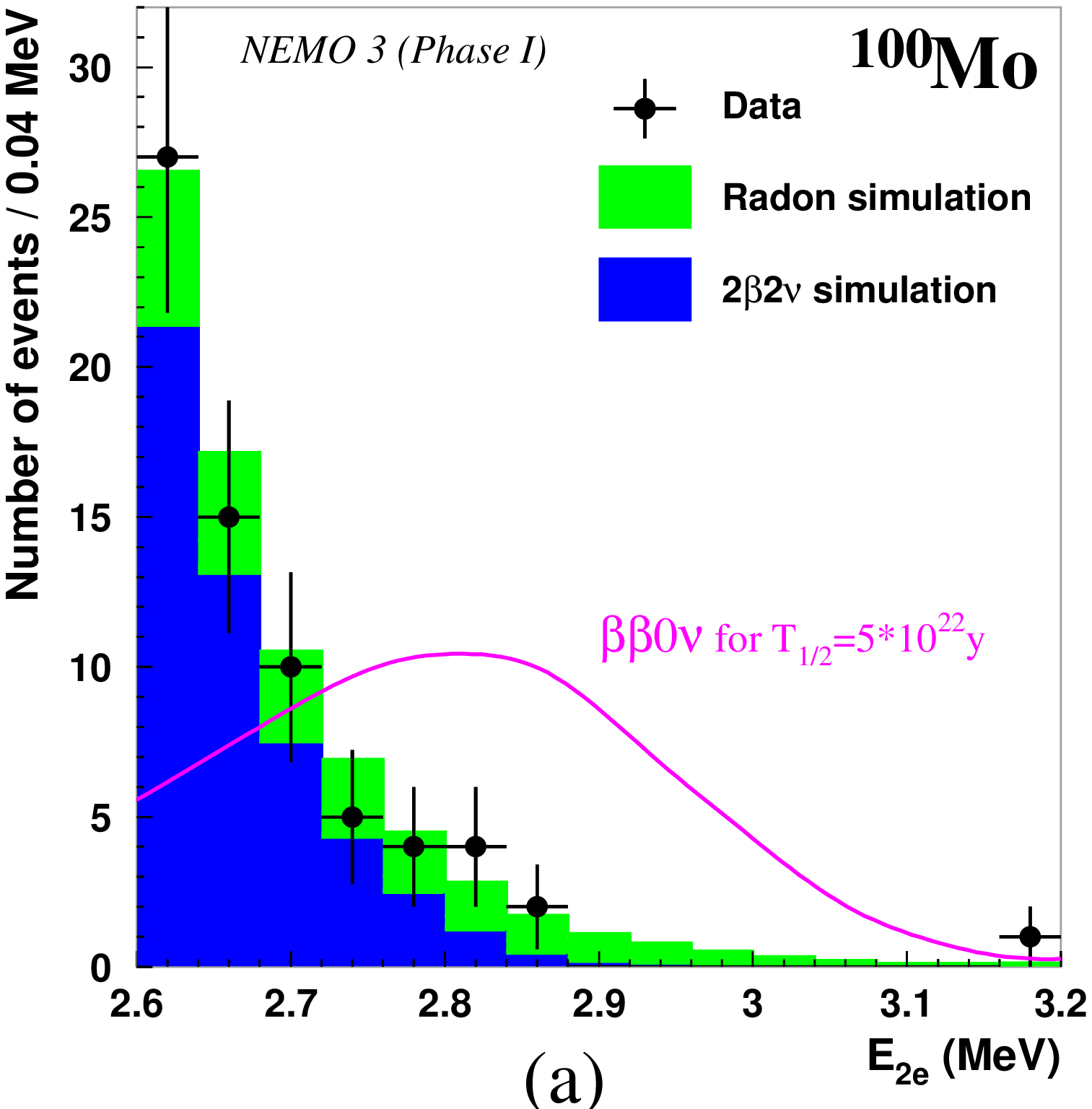}
\includegraphics[scale=0.2]{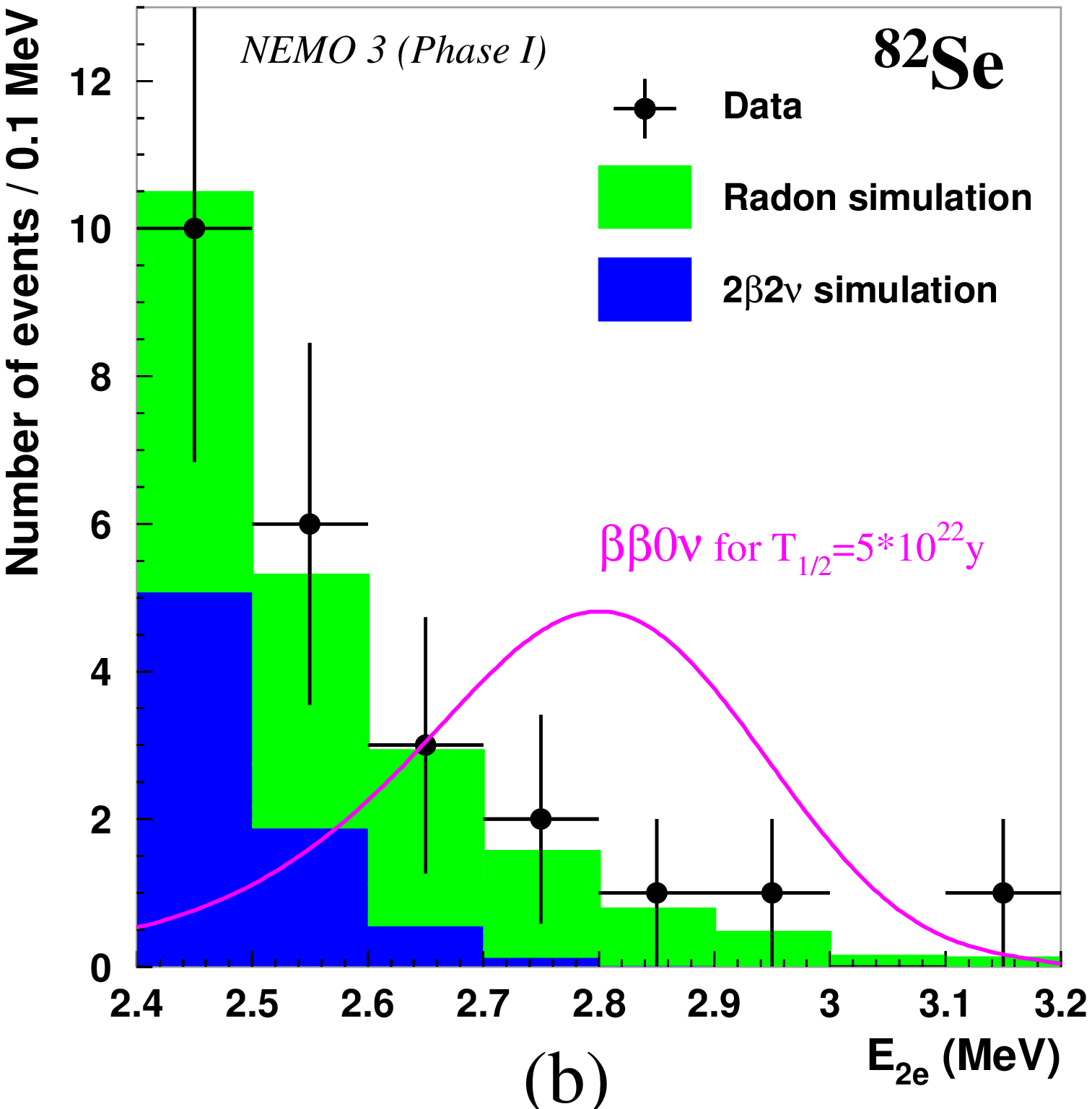}
\includegraphics[scale=0.2]{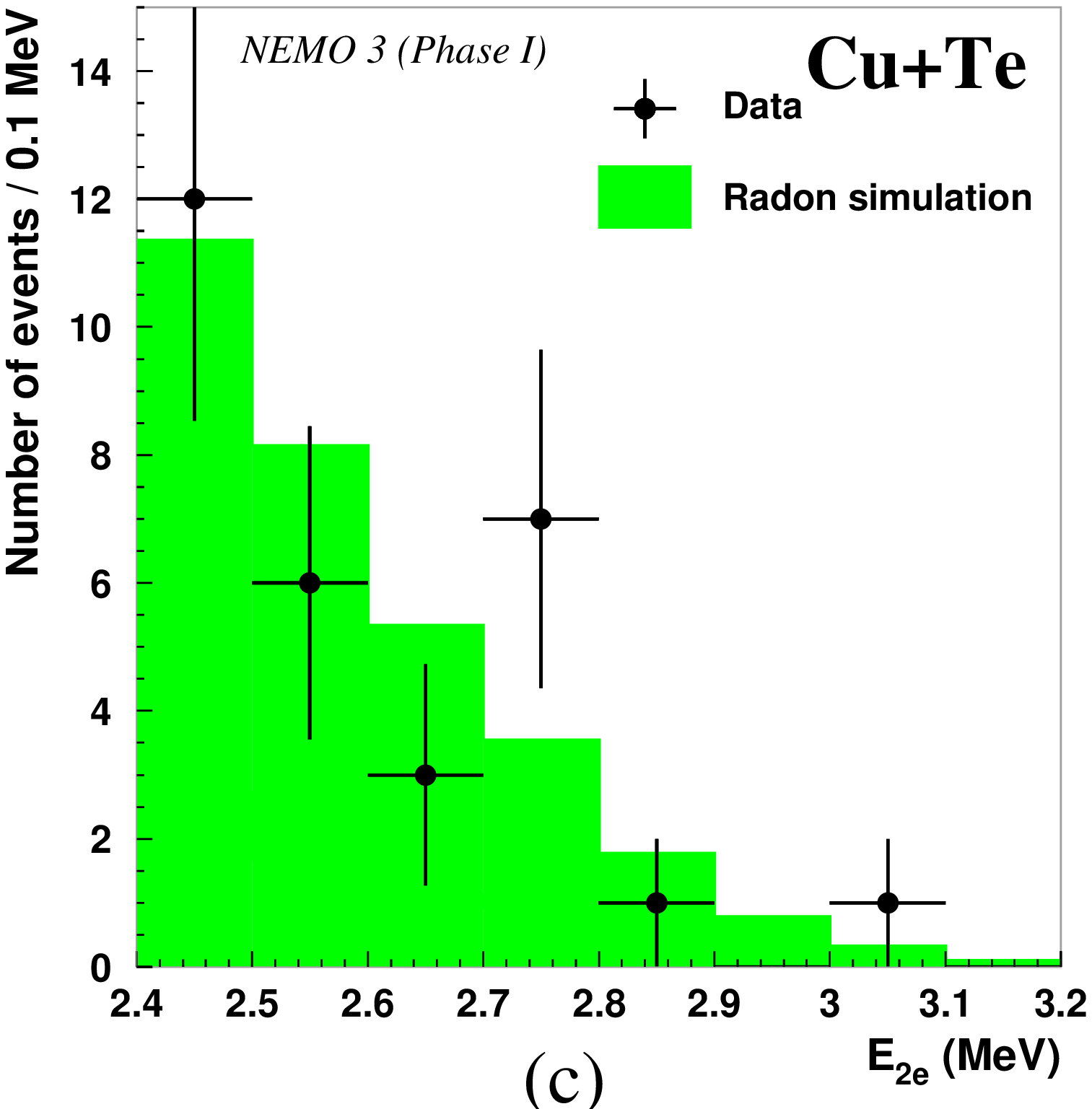}
\caption{\label{fig:fugure6} Spectra of the energy sum of the two
electrons in the $2\beta(0\nu)$ energy window
after 389 effective days of data collection from February 2003
until September 2004 (Phase~I):
(a) with  $6.914$~kg of $^{100}$Mo; (b) with $0.932$~kg of
$^{82}$Se; (c) with Copper and Tellurium foils \cite{ARN05}.
The shaded histograms are the expected backgrounds computed by
Monte-Carlo simulations:
dark (blue) is the $2\beta(2\nu)$ contribution and light (green)
is the Radon contribution.
The solid line corresponds to the expected $2\beta(0\nu)$
signal if $T_{1/2} = 5 \times 10^{22}$ y.}
\end{center}
\end{figure*}

\subsection{CUORICINO \cite{ARNA05}}

This program is the first stage of the larger CUORE experiment
(see Subsection 4.1). The experiment
is running at the Gran Sasso Underground Laboratory in Italy 
(3500 m w.e.). The detector
consists of low-temperature devices
based on $^{nat}$TeO$_{2}$  crystals. The use of natural tellurium is
justified in this case, because the content of
$^{130}$Te
in it is rather high, 33.8\%. The detector consists of 62
individual crystals, their total weight being 40.7 kg.
The energy resolution is 7.5-9.6 keV at an energy of 2.6 MeV.

\begin{figure}
\begin{center}
\includegraphics[scale=0.4]{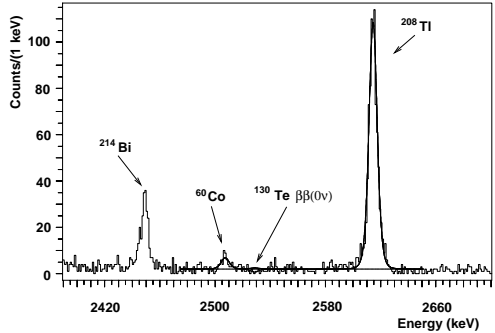}
\caption{\label{fig:figure7} The sum spectra of all crystal in the
region
of the $2\beta(0\nu)$ energy \cite{ARNA05}.}
\end{center}
\end{figure}

The experiment has been running
 since March of 2003. The summed spectra of all crystals in the
region
of the $2\beta(0\nu)$ energy is shown in Fig.~\ref{fig:figure7}.
The total exposure is 3.09 $kg \cdot$ y  ($^{130}$Te). The
background
at the energy of the $2\beta(0\nu)$ decay is 0.18 keV$^{-1}\cdot
kg^{-1} \cdot y^{-1}$.  No peak is evident and the limit is
$T_{1/2} > 1.8\cdot 10^{24}$ y (90\% C.L.)\footnote{It should be stressed that 
"sensitivity" of the experiment under present conditions (when number 
of observed events is equal to expected mean background) is $\sim 
1\cdot 10^{24}$ y (90\% C.L.). Much better limit was obtained due to 
big "negative" fluctuation of the background in the $0\nu$ energy region.}.

Using NME values from \cite{SIM99,STO01,CIV03} the limit on
$\langle m_{\nu} \rangle$ is less than 0.4-0.9 eV.
If one uses the NME from \cite{ROD05}
then $\langle m_{\nu} \rangle < 1.1-1.6$ eV.

The sensitivity of the experiment to $2\beta(0\nu)$ decay of
$^{130}$Te under present conditions will be on the level 
of $\sim 4\cdot 10^{24}$ at 90\%C.L. for
5 y of measurement. This in turn means the sensitivity to $\langle
m_{\nu} \rangle$  is on the level of 0.3-0.9 eV. At the same time 
there is a hope to detect $2\beta(2\nu)$ decay of $^{130}$Te
in this experiment.

One of the tasks of the CUORICINO experiment is to demonstrate the
possibility of
substantially reducing of the background to the level of 0.01-
0.001 keV$^{-1}\cdot kg^{-1} \cdot y^{-1}$
which is necessary to proceed with the
 CUORE Project (see section 4.1).

\section{Planned experiments}

In this section, mention of five most developed and 
promising experiments which can
be realized within the next five to ten years is discussed 
(see Table 7). The
estimation of the sensitivity in all experiments is made using
NMEs
from \cite{SIM99,STO01,CIV03,ROD05}.

\begin{table}[ht]
\label{Table7}
\caption{Five most developed and promising projects (see text). 
Sensitivity at 90\% C.L. for three (1-st step of GERDA and MAJORANA) 
five (EXO, SuperNEMO) and ten (CUORE, full-scale GERDA and MAJORANA) 
years of measurements is presented. 
[$^{*)}$ For the background 
0.001 keV$^{-1}\cdot kg^{-1} \cdot y^{-1}$; $^{**)}$ for the background 
0.01 keV$^{-1}\cdot kg^{-1} \cdot y^{-1}$.] }
\vspace{0.5cm}
\begin{center}
\begin{tabular}{cccccc}
\hline
Experiment & Isotope & Mass of & Sensitivity  & Sensitivity & Status \\
& & isotope, kg & $T_{1/2}$, y & $\langle m_{\nu} \rangle$, meV &  \\
\hline
CUORE \cite{ARNA04} & $^{130}$Te & 200 & $4.6\cdot10^{26}$$^{*)}$ & 30-100 & accepted \\ 
& & & $1.4\cdot10^{26}$$^{**)}$ & 40-170 & \\
GERDA \cite{ABT04} & $^{76}$Ge & 40 & $2\cdot10^{26}$ & 90-300 & accepted \\
& & 500 & $4\cdot10^{27}$ & 20-70 & R\&D\\ 
MAJORANA & $^{76}$Ge & 180 & $5\cdot10^{26}$ & 60-200 & R\&D \\
\cite{MAJ03, AAL05}& & 500 & $4\cdot10^{27}$ & 20-70 & R\&D \\ 
EXO \cite{DAN00} & $^{130}$Xe & 200 & $6.4\cdot10^{25}$ & 70-400 & accepted \\
& & 1000 & $8\cdot10^{26}$ & 12-86 & R\&D \\ 
SuperNEMO & $^{82}$Se & 100 & $(1-2)\cdot10^{26}$ & 40-150 & R\&D \\
\cite{BAR02,BAR04a,PIQ05} & & & & &\\
\hline
\end{tabular}
\end{center}
\end{table}

\subsection{CUORE \cite{ARNA04}}

This experiment is to be
run at the Gran Sasso Underground Laboratory (Italy; 3500 m w.e.). 
The plan is to
investigate 760 kg of $^{nat}$TeO$_{2}$ , with a total of
206 kg of $^{130}$Te. One thousand low-temperature ($\sim$ 8 mK)
detectors, each having a weight of 760 g, will be
manufactured and arranged in 25 towers (one tower is approximately
equivalent to the CUORICINO detector, see
Subsection 3.2). Planed energy resolution is 5 kev (FWHM).
 One of the main problems here is to reduce the
background level by a factor of about 10
to 100 in relation to the background level achieved in the
detector CUORICINO \cite{ARNA05}. Upon reaching a
background level of 0.001 keV$^{-1}\cdot kg^{-1} \cdot y^{-1}$,
the sensitivity of the experiment to
the $0\nu$ decay of $^{130}$Te for 10 y of measurements and at
90\% C.L. will
become approximately
$4.6 \cdot 10^{26}$ y ($\langle m_{\nu} \rangle$ $\sim$ 0.03-0.1
eV). For more realistic level of background  
0.01 keV$^{-1}\cdot kg^{-1} \cdot y^{-1}$ 
sensitivity will be $\sim 1.4\cdot 10^{26}$ y for half-life and 
$\sim$ 0.04-0.17 eV for effective Majorana neutrino mass. 

The experiment has been approved and funded.

\subsection{GERDA \cite{ABT04}}

This is one of two (along with the MAJORANA experiment) planned
experiments
with $^{76}$Ge. The experiment is to be
located in Gran Sasso Underground Laboratory (3500 m w.e.). The
proposal is based on ideas and approaches which
were proposed for GENIUS \cite{KLA98} and the GEM \cite{ZDE01}
experiments.
The plan is to place "naked" HPGe detectors in highly
purified liquid nitrogen. It minimizes the weight of construction
material near the detectors and, as a result,
decreases the level of background. The liquid nitrogen dewar is
placed into a vessel of very pure water.
The water plays a role of passive and active (Cherenkov radiation)
shield.

The proposal involves three phases. In the first phase, the
existing HPGe detectors ($\sim$ 15 kg), which
previously were used in Heidelberg-Moscow \cite{KLA01} and IGEX
\cite{AAL02} experiments, will be utilized. In the second phase
$\sim$ 40 kg of enriched Ge will be investigated. In the third
phase the plan
 is to use $\sim$ 500 kg of $^{76}$Ge.

The first phase, lasting one year, is to measure the sensitivity to
$3 \cdot 10^{25}$ y, which gives a possibility
to checking the "positive" result of \cite{KLA04}. The
sensitivity of the second phase (for three years of measurement)
will be $\sim$ $2 \cdot 10^{26}$ y, which corresponds to a
sensitivity for $\langle m_{\nu} \rangle$ on the level of
$\sim$ 0.09-0.3 eV.

The first two phases have been approved and funded. Measurements 
will start in $\sim$
2007(2008). The results of this
first step will play an important role in the decision to support
the full scale experiment.

The project is very promising although it will be difficult to
reach the desired level of background.
One of the significant problems is $^{222}$Rn in the liquid
nitrogen (see, for example, results of \cite{KLAP04}).

\subsection{MAJORANA \cite{MAJ03, AAL05}}

The MAJORANA facility will consist of 210 sectioned HPGe detectors
manufactured from enriched germanium (the degree
of enrichment is about 86\%). The total mass of enriched germanium
will be 500 kg. The facility is designed in such a
way that it will consist of ten individual supercryostats
manufactured from low radioactive copper, each containing
21 HPGe detectors. The entire facility will be surrounded by a
passive shield and will be located at an underground
laboratory in Canada (or in the United States).
Only the total energy deposition will be utilized in measuring the
$2\beta(0\nu)$ decay of $^{76}$Ge to the ground
state 
of the daughter nucleus. The use of sectioned HPGe detectors,
pulse shape analysis, anticoincidence,
and low radioactivity structural materials will make it possible
to reduce
the background to a value below $3 \cdot 10^{-4}$ keV$^{-1}\cdot
kg^{-1} \cdot y^{-1}$  and to reach
a sensitivity of about $4 \cdot 10^{27}$ y within
ten years of measurements. The corresponding sensitivity to the
effective mass of the Majorana neutrino is about
0.02 to 0.07 eV.
The measurement of the $2\beta(0\nu)$ decay of $^{76}$Ge to the
0$^{+}$ excited state of the daughter nucleus
will be performed
by recording two cascade photons and two beta electrons. The
planned sensitivity for this process is about 10$^{27}$ y.

In the first step $\sim$ 180 kg of $^{76}$Ge will be
investigated. It is anticipated that in the sensitivity to
$2\beta(0\nu)$ decay to the ground state of the daughter nuclei
for 3 years of measurement will be $5\cdot 10^{26}$ y.
It will reject or to confirm the "positive" result from
\cite{KLA04}. Sensitivity to
$\langle m_{\nu} \rangle$ will be $\sim$ 0.06-0.2 eV. During this
time different methods and technical
questions will be checked and possible background problems will be
investigated.

\subsection{EXO \cite{DAN00}}

In this experiment the plan is to implement M. Moe's proposal of
1991 \cite{MOE91}. Specifically it is to record both ionization
electrons
and the Ba$^{+}$ ion originating from the double-beta-decay
process $^{136}$Xe
-
$^{136}Ba^{++}$ + 2e$^{-}$. In reference \cite{DAN00}, it is
proposed to
operate with 1t of $^{136}$Xe. The actual technical implementation
of
the experiment has not yet been
developed. One of the possible schemes is to fill a TPC
with liquid enriched xenon. To avoid the background from the
2$\nu$ decay of $^{136}$Xe, the
energy resolution
of the detector must not be poorer than 3.8\% (FWHM) at an energy
of 2.5 MeV (ionization and scintillation signals
will be detected).

In the 0$\nu$ decay of $^{136}$Xe, the TPC will measure the energy
of two electrons and the coordinates of the event
to within a few millimeters. After that, 
using special stick Ba ion will be removed from the liquid and then 
will be registered in special cell 
by resonance excitation. For Ba$^{++}$ to undergo
a transition to a state of Ba$^{+}$, a special gas is added to
xenon. The authors of the project assume that the
background will be reduced to one event within five years of
measurements. Given 70\% detection efficiency
it will be possible to reach a sensitivity of about $8 \cdot
10^{26}$ y for the $^{136}$Xe half-life and a
sensitivity of about 0.012 to 0.086 eV for the neutrino mass.

The authors also considered a detector in which the mass of
$^{136}$Xe is 10 t, but this is probably
beyond present-day capabilities. It should be noted that about 100
t of natural xenon are required to obtain
10 t of $^{136}$Xe.  This exceeds the xenon produced worldwide
over several years.

One should note that the principle difficulty in this experiment
is associated with detecting the
 Ba$^{+}$ ion with a reasonably high efficiency under
 conditions of real experiment.
This issue calls for thorough experimental tests, and positive
results along these lines have yet to be obtained.

As the first stage of the experiment it is planned the EXO-200
will use 200 kg of $^{136}$Xe without Ba ion registration.
This experiment is currently
under preparation and measurement will start probably in
2007(2008).
The 200 kg of enriched Xe is a product of
Russia (enrichment is $\sim 80\%$). If the background will be 
40 events within 5 y of measurements,
 as estimated by the
authors, then the sensitivity of the experiment will be $\sim$
$6 \cdot 10^{25}$ y (this corresponds to sensitivity for $\langle
m_{\nu} \rangle$ at the level $\sim$ 0.07-0.4 eV). This initial prototype 
will operate at the Waste Isolation Pilot Plant (WIPP) in Southern 
New Mexico (USA).

\subsection{SuperNEMO \cite{BAR02,BAR04a,PIQ05}}

The NEMO Collaboration has studied and is pursing an experiment
that will observe 100 kg of $^{82}$Se
with the aim of reaching a sensitivity to the $0\nu$ decay mode at
the level of
$T_{1/2} \sim (1-2) \cdot10^{26}$ y
(the corresponding sensitivity to the neutrino mass is about 0.04
to 0.15 eV). In order to accomplish this
goal, it is proposed to use the experimental procedures nearly
identical to that in the NEMO-3 experiment
(see Subsection 3.1). The new detector will have planar geometry
and will consist of 20
identical modules (5 kg of $^{82}$Se in each sector). A $^{82}$Se
source having a thickness of
about 40 mg/cm$^{2}$ and a very low content of radioactive
admixtures is placed at the center of
the modules. The detector will again record all features of double
beta
decay: the electron energy will be recorded by counters based on
plastic scintillators ($\Delta E/E \sim 10-12\% (FWHM) $
at E = 1 MeV),
while tracks will be reconstructed with the aid of Geiger
counters.
The same device can be used to investigate $^{100}$Mo, $^{116}$Cd,
and $^{130}$Te with a sensitivity to
$2\beta(0\nu)$ decay at a
level of about $(0.5-1) \cdot$ 10$^{26}$ y.

The use of an already tested experimental technique is an
appealing feature of this experiment. The plan is
to arrange the equipment at the new Frejus Underground Laboratory
(France; the respective depth being 4800 m w.e.) or at CANFRANC Underground 
Laboratory (Spain; 2500 m w.e.).
The experiment is currently in its  R\&D stage.

\section{Conclusion}

In conclusion, two-neutrino double-beta decay has so far been
recorded for ten nuclei ($^{48}$Ca, $^{76}$Ge,
$^{82}$Se, $^{96}$Zr, $^{100}$Mo,
$^{116}$Cd, $^{128}$Te, $^{130}$Te, $^{150}$Nd, $^{238}$U). In
addition, the $2\beta(2\nu)$ decay of $^{100}$Mo
 and $^{150}$Nd to 0$^{+}$ excited state of the daughter
nucleus has been observed and the ECEC(2$\nu$) process in
$^{130}$Ba was recorded. Experiments studying two-neutrino double
beta decay are presently approaching a qualitatively new level,
where high-precision measurements are performed
not only for half-lives but also for
all other parameters of the process.
As a result, a trend is emerging toward thoroughly investigating
all aspects of two-neutrino double-beta decay,
and this will furnish very important information about the values
of nuclear matrix elements, the parameters
of various theoretical models, and so on. In this connection, one
may expect advances in the calculation of
nuclear matrix elements and in the understanding of the nuclear-
physics aspects of double beta decay.

Neutrinoless double beta decay has not yet been confirmed. There
is
a conservative limit on the effective value
of the Majorana neutrino mass at the level of 0.9 eV. Within the
next few years,
the sensitivity to the neutrino mass in the CUORICINO and NEMO-3
experiments will be improved to become about
0.3 to 0.9 eV with measurements of $^{130}$Te and $^{100}$Mo. With
the NEMO-3 detector, a similar level of sensitivity
can be reached for some other nuclei as well (with 10 kg of
$^{82}$Se, for example). It is precisely these two experiments
(NEMO-3 and CUORICINO) that will carry out the investigations of
double beta decay over the next three to five years.
The Next-generation experiments, where the mass of the isotopes
being studied
will be as grand as 100 to 1000 kg, will have started within five
to ten years. In all probability,
they will make it possible to reach the sensitivity to the
neutrino mass at a level of 0.1 to 0.01 eV.

\section*{Acknowledgments}

I would like to thank Prof. C.S. Sutton for his useful remarks.
A portion of this work was supported by grants from INTAS (03051-
3431) and NATO (PST CLG 980022).


\end{document}